\documentclass[journal,twoside,web]{ieeecolor}
\usepackage{jsen}
\usepackage{cite}
\usepackage{amsmath,amssymb,amsfonts}
\usepackage{algorithmic}
\usepackage{graphicx}
\usepackage{textcomp}
\usepackage{wrapfig}
\usepackage{orcidlink}
\usepackage{titlesec}
\usepackage{comment}
\usepackage{tikz}
\usepackage{gensymb}
\usetikzlibrary{shapes.geometric, arrows}
\tikzstyle{startstop} = [rectangle, rounded corners, minimum width=3cm, minimum height=1cm,text centered, draw=black, fill=red!30]
\tikzstyle{process} = [rectangle, minimum width=3cm, minimum height=1cm, text centered, draw=black, fill=orange!30]
\tikzstyle{arrow} = [thick,->,>=stealth]

\def\BibTeX{{\rm B\kern-.05em{\sc i\kern-.025em b}\kern-.08em
    T\kern-.1667em\lower.7ex\hbox{E}\kern-.125emX}}
\definecolor{abstractbg}{RGB}{255,255,255}

\definecolor{ssred}{RGB}{230,193,194} % Define ssred
\definecolor{ssblue}{RGB}{194,213,233} % Define ssblue

\begin{document}
\title{Cultivating Precision: Comparative Analysis of Sensor-Based Yogurt Fermentation Monitoring Techniques}

\author{Ege Keskin\orcidlink{0000-0002-5684-2229}, İhsan Ozan Yıldırım\orcidlink{0000-0002-2432-147X}
%\thanks{Manuscript received 2 May 2018; revised 9 September 2018; accepted 12 October 2018. Date of publication 29 November 2018; date of current version 7 March 2019. \textit{(Ege Keskin and İhsan Ozan Yıldırım are co-first authors) (Corresponding authors: E. Keskin; İ.O. Yıldırım)}}
% \thanks{This paragraph of the first footnote will contain the date on which you submitted your paper for review. It will also contain support information, including sponsor and financial support acknowledgment. For example, ``This work was supported in part by the U.S. Department of Commerce under Grant BS123456.'' }
% \thanks{The next few paragraphs should contain the authors' current affiliations, including current address and e-mail. For example, F. A. Author is with the National Institute of Standards and Technology, Boulder, CO 80305 USA (e-mail: author@boulder.nist.gov). }
%\thanks{S. B. Author, Jr., was with Rice University, Houston, TX 77005 USA. He is now with the Department of Physics, Colorado State University, Fort Collins, CO 80523 USA (e-mail: author@lamar.colostate.edu).}
\thanks{E. Keskin and İ. O. Yıldırım are with Beko Corporate, R\&D Directorate, Sensor Technologies, İstanbul 34912, Türkiye (e-mail: \href{mailto:ege.keskin@arcelik.com}{ege.keskin@beko.com}, \href{mailto:ihsanozan.yildirim@arcelik.com}{ihsanozan.yildirim@beko.com})}
% If an article is a thesis or part of a thesis or dissertation, this should be so noted in the last sentence of the first paragraph of the footnote.
%\thanks{}
}

\IEEEtitleabstractindextext{%
\fcolorbox{abstractbg}{abstractbg}{%
\begin{minipage}{\textwidth}%https://www.overleaf.com/project/63d3a5971c5182e3849fc617
\begin{wrapfigure}{r}{0.45\textwidth}%
\includegraphics[width=0.43\textwidth]{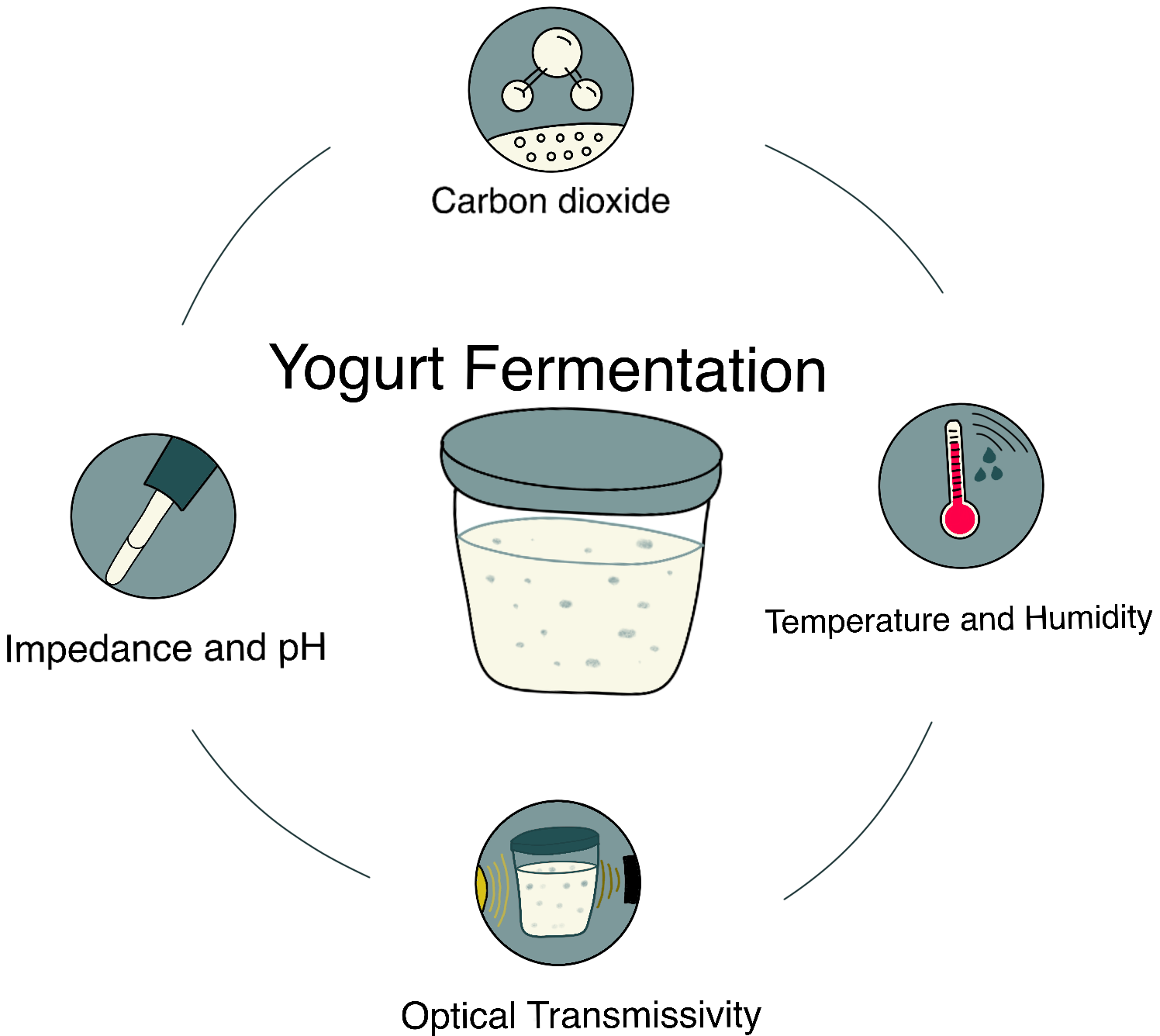}%
\centering
\end{wrapfigure}%

\begin{abstract}

Fermented dairy products, including yogurt, are widely consumed for their nutritional and health benefits. While numerous methods exist to monitor and understand yogurt fermentation, the literature lacks an integrated evaluation of diverse sensing approaches within a single experimental framework. To address this gap, this study systematically examines and compares multiple measurement techniques—electrical impedance, DC resistance, pH, optical transparency, carbon dioxide concentration, ambient temperature, and relative humidity—in tracking the yogurt fermentation process. By presenting a unified set of experimental results and assessing each method’s observational characteristics, this work offers an encompassing reference point for researchers seeking to understand the relative merits and limitations of different sensing modalities. Rather than establishing definitive guidelines or practical recommendations, the findings provide a foundation for subsequent investigations into sensor-based fermentation monitoring, thereby contributing to a more comprehensive understanding of yogurt fermentation dynamics.

\end{abstract}

\end{minipage}}}

\maketitle

%%%%%%%%%%%%%%%%%%%
% OLMASI GEREKENLER:
% - Abstractta:  İlk kez siz ne katkıda bulundunuz?
% - Girişte: Bu soru kıymetli mi:  Siz bunun için ne bekliyorsunuz?
% - Sonuçta: Yaptınız da ne oldu, oldu da ne oldu? 
% - Sonuçta: Bu sonuç ne kadar güvenilir?
% - Sonuçta: Bundan sonrası için ne söylüyorsunuz?
% - Sonuçta: İlk kez söyleoghurt-jsen-diğiniz şey ne, katkı, contribution ne?
%%%%%%%%%%%%%%%%%%%

\section{Introduction}

Yogurt, a fermented dairy product, has seen a significant increase in demand in recent years, driven by its nutritional profile, health benefits, and wide range of available flavors \cite{shahbook}. Multiple studies have underscored advantages such as reduced lactose intolerance and improvements in gut health \cite{CHANDAN201731}. Correspondingly, a growing number of consumers have ventured into preparing yogurt at home \cite{yogurthome}. While the basic production process is relatively simple, several factors—such as difficulty ensuring preservative-free milk, limited knowledge, time constraints, and inconsistent outcomes—often discourage these efforts.

The fermentation process is inherently sensitive to various parameters, including the milk’s composition (fat, protein, sugar, and other constituents), the bacterial culture strains and their activity, as well as the fermentation temperature and humidity regime. This complexity makes it challenging to pinpoint the exact moment at which the desired taste and texture are achieved. An overly prolonged process may yield excessively sour yogurt, whereas premature termination can result in underdeveloped flavors and inadequate consistency.

In principle, the integration of appropriate sensing technologies could enable automated systems to detect the fermentation state with greater precision. Such a setup could terminate heating and begin cooling at the ideal juncture, potentially alerting users through simple notifications or via connectivity with smart home platforms like HomeWhiz \cite{HomeWhiz}, Google Home \cite{GoogleHome}, or similar applications. These approaches may enhance both the consistency and user experience of homemade yogurt fermentation.

Yogurt is an ideal subject for this investigation due to its broad popularity and recognized health benefits. Consumers gravitate towards homemade production out of concern for commercial additives or preservatives \cite{yogurthome}, and among families with newborns there is a pronounced interest in avoiding unfamiliar chemical inputs. As health consciousness rises, so does the desire for safer, more transparent production methods, whether using traditional techniques or dedicated yogurt-making devices. However, effective and accessible methods to monitor the fermentation process in a domestic setting remain limited.

Despite the potential utility of such measurements, there are currently no widely available household products or established methods for accurately tracking yogurt fermentation progress. To address this gap, the present study systematically investigates and compares various monitoring techniques. It aims to provide a consolidated examination of measurement methods—evaluating their capabilities, as well as their respective strengths and limitations—without prescribing specific best practices or recommendations.

The comprehensive overview offered here is pertinent to a broad audience. Individuals and researchers interested in sensing technologies, data processing, microcontrollers, communication protocols, and sensor integration can benefit from the findings. The work is also relevant to engineers and professionals in fields such as control systems, mechatronics, electrical and computer engineering, as well as to product developers and food technologists. By examining and comparing a range of sensing modalities, this study serves as a detailed resource for understanding the state-of-the-art in yogurt fermentation monitoring, offering a knowledge foundation upon which future investigations and innovations can build.

\section{Background and Related Work} 

A thorough understanding of existing and established approaches in the literature is essential for this study, as most of the current measurement methods may assist in addressing our research objectives. This section is organized into two parts: first, an overview of the yogurt fermentation process, and second, a review of existing sensing technologies for dairy products.

\subsection{Yogurt Fermentation}

According to the Oxford English Dictionary, fermentation is defined as "the chemical breakdown of a substance by bacteria, yeasts, or other microorganisms, typically involving effervescence and the giving off of heat." Yogurt is produced by fermenting milk with specific bacterial cultures, primarily \emph{Lactobacillus delbrueckii} subsp. \emph{bulgaricus} and \emph{Streptococcus thermophilus} \cite{shahbook}. These thermophilic organisms have optimal growth rates at approximately 45°C, with a maximum temperature of 50°C. They are usually mixed in a 1:1 ratio and function symbiotically to develop the characteristic properties of yogurt. However, neither of these cultures alone is capable of producing the optimal balance of acid and flavor.

\emph{S. thermophilus} initiates lactic acid production and lowers the oxygen concentration, which stimulates the growth of \emph{L. delbrueckii} subsp. \emph{bulgaricus} \cite{yogurttech}. As a result of this symbiotic relationship, lactic acid is produced, imparting yogurt its sourness and acidity. Lactic acid fermentation is thus a crucial chemical reaction in yogurt formation.

The average fermentation time for yogurt ranges from 4 to 6 hours, depending on factors such as the initial concentration of bacterial cultures, fermentation temperature, and the desired final acidity and texture. During this period, the bacterial cultures metabolize lactose into lactic acid, leading to a decrease in pH and causing the milk proteins to coagulate, which contributes to the yogurt's characteristic texture and flavor.

As a final step, the yogurt is typically chilled to 4°C for at least 8 hours to ensure solid coagulation and to halt further fermentation \cite{cooling}. This cooling process stabilizes the yogurt's structure and extends its shelf life by slowing down bacterial activity.

\subsection{Measurement Methods for Dairy Products}

Milk exhibits both resistive and capacitive properties due to its composition and the interactions between measurement probes and the milk itself. Consequently, the electrochemical measurement model of milk can be represented as a resistor and two capacitors connected in series when two probes are used \cite{impedanceecoli}.

Electrical Impedance Spectroscopy (EIS) is a technique wherein the electrical impedance of a medium or test subject is measured across a range of frequencies. Certain materials exhibit unique properties at specific frequencies, allowing for a more detailed analysis of the material \cite{eis}. Previous work has demonstrated that EIS can be used to detect the coagulation point of yogurt and assess its acidity \cite{impedanceyoghurt}. The motivation behind using EIS for monitoring yogurt fermentation lies in its ability to provide a sanitary and non-destructive measurement method.

As previously mentioned, lactic acid production during yogurt preparation results in a decrease in the pH of the mixture, increasing its acidity. The most direct method of measuring the acidity or alkalinity of a substance is to use a pH probe, where pH stands for "potential of hydrogen." A pH probe consists of a glass electrode, a reference electrode, and a reference solution, which together convert the difference in hydrogen ion concentration between the medium and the reference solution into a voltage signal \cite{phbook}. This voltage signal can then be measured and converted to the actual pH value using established equations. Although pH probes provide a direct measurement of acidity or alkalinity, they require maintenance and calibration to ensure stable operation \cite{phcalibration}. This calibration procedure often includes different reference liquids with specific pH levels and can be cumbersome for the ordinary user.

Optical methods are often preferred for the analysis of dairy products because they do not require direct contact with the sample. It has been reported that the fat content of milk affects its light permeability \cite{opticalfat}. This phenomenon was demonstrated using a stable light source and a light-dependent resistor (LDR); as the fat content of the milk varied, the voltage across the LDR changed due to variations in transmitted light. Similarly, laser diffraction patterns have been used to classify different types of milk \cite{laserdiffraction}.

\subsection{Existing Sensor Systems and Benchmark Products}

Recent advancements have led to the development of various sensor systems and devices aimed at monitoring fermentation processes, particularly in industrial applications \cite{industrialYogurt}. In the dairy industry, large-scale fermentation tanks are often equipped with sensors that measure parameters such as temperature, pH, dissolved oxygen, and microbial activity. These sensors enable precise control over the fermentation environment, ensuring product consistency and quality.

For consumer-level applications, several commercial yogurt makers are available that automate the fermentation process to some extent. These devices typically provide temperature control and timers to maintain the optimal conditions for bacterial growth. However, they generally lack advanced sensing capabilities to monitor the fermentation progress in real time. Users must rely on preset fermentation times, which may not account for variations in milk composition, ambient conditions, or personal taste preferences.

Some innovative consumer devices have attempted to incorporate basic sensors. For example, certain yogurt makers include built-in thermometers to display the internal temperature, allowing users to ensure that the device maintains the correct fermentation temperature. However, these are limited to temperature sensing and do not provide information on other critical parameters like pH or acidity.

In research settings, various sensor technologies have been explored for fermentation monitoring. Electrical impedance spectroscopy has been used to assess microbial growth and fermentation kinetics, providing insights into the progress of fermentation without direct sampling \cite{eisYogurt}. Optical sensors, such as near-infrared spectroscopy, have also been investigated for their ability to monitor changes in the fermentation medium by detecting variations in light absorption and scattering due to microbial activity \cite{nearIR}.

Despite these technological advancements, there remains a lack of comparison regarding these measurement methods with systematic experimentation. 

Reviewing previous work allows us to understand the microbiological and chemical processes involved in yogurt fermentation, as well as the measurement methods used for assessing product quality and monitoring the fermentation process. We will synthesize this information to develop a testing plan for potential sensor applications.

\section{Method and Experimental Design}

We formulated our methodology around a problem statement and a research question to draw the boundaries of the study. Based on the hypothesis that certain measurable properties of yogurt correlate with its pH, we sought to develop a list of potentially cost-effective and accurate sensor options track fermentation, allowing retrieval of the product at the desired acidity level.

\subsection{Method}

The research was structured around the following key components:

\begin{itemize} 
\item \textbf{Problem Statement}: Observing the yogurt fermentation process is not thoroughly explored with an array of different sensor solutions. Hence, an organized comparison of sensing methods is lacking in the literature.
\item \textbf{Hypothesis}: Directly or indirectly tracking the pH of yogurt during fermentation can provide an accurate indication of the fermentation stage and desired acidity level. 
\item \textbf{Research Question}: Which physical and chemical properties of yogurt are correlated with its pH during fermentation? 
\item \textbf{Approach}: Investigate various measurement methods to discuss their strengths and weaknesses, ultimately resulting in an even comparison of these methods in terms of performance. 
\end{itemize}

To explore the correlation between yogurt's pH and other measurable properties, we constructed various test setups with different volumes, geometries, and specifications. Consistent combinations of milk and bacterial culture were used across all fermentation runs to minimize uncontrolled variables. Each sensor system was tested systematically, and the results were analyzed for their relevance and correlation to acidity.

\subsection{Materials}

For the experiments, the following materials and equipment were utilized:

\begin{itemize} 
\item \textbf{Milk and Culture}: Commercial whole milk and a standardized yogurt culture were used for all fermentation runs to ensure consistency, as they are a direct input to the fermentation trend and pH change profile.
\item \textbf{Fermentation Cavity}: Re-purposed containers equipped with various sensors to collect data during fermentation. 
\item \textbf{Heat Source}: A household oven with a yogurt making function was used to provide a stable heat source to the milk and culture mixture.

\item \textbf{Sensors and Measurement Systems}: 
\begin{itemize} 
\item CM1106 Wuhan Cubic CO\textsubscript{2} sensor (positioned on the lid at 3 o'clock) 
\item SHT31 Sensirion ambient temperature and humidity sensor (positioned on the lid at 6 o'clock) 
\item Logitech Webcam for surface imaging (located at the center of the lid) 
\item Electrical probes for impedance and DC resistance measurements 
\item Thermocouples for measuring the temperature of the yogurt mixture 
\item Si1133 Silicon Labs ambient light sensor 
\end{itemize} 
\item \textbf{Data Acquisition System}: Raspberry Pi 4 for data collection and sensor communication master device. 
\item \textbf{Software}: Python scripts developed for sensor interfacing and data logging. 
\end{itemize}

\begin{figure}[h] 
\centering 
\includegraphics[width=0.45\textwidth]{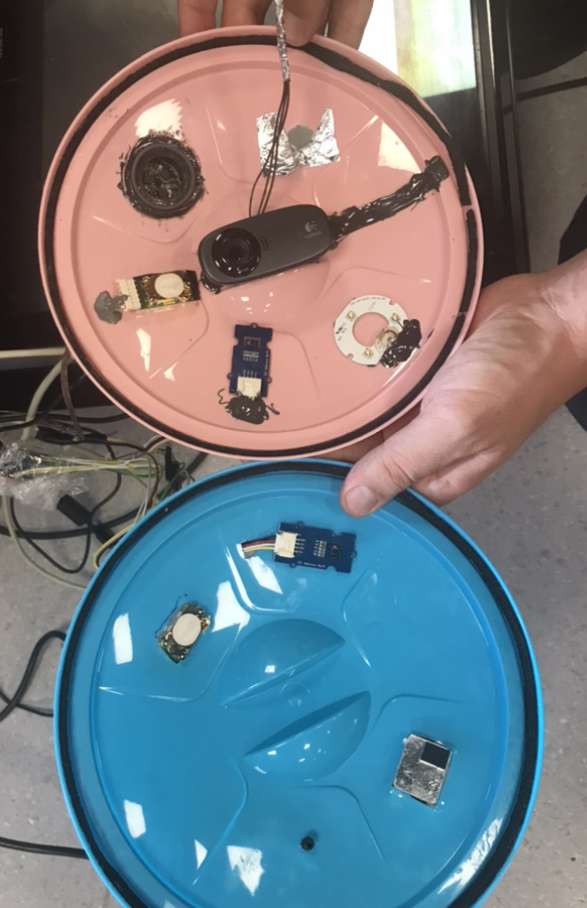} 
\caption{Measurement setup with various sensors integrated into the fermentation container lid. The seam of the lid was covered with a soft tape to provide thermal insulation.} 
\label{figLid} 
\end{figure}

\begin{figure}[h] 
\centering 
\includegraphics[width=0.45\textwidth]{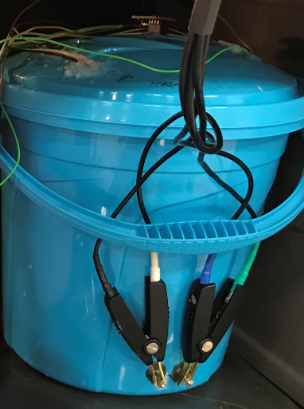} 
\caption{Electrical impedance measurement contacts inside the fermentation container.} 
\label{ImpProbes} 
\end{figure}

Figures \ref{figLid} and \ref{ImpProbes} show the measurement setup and the placement of sensors within the fermentation container.

\subsection{Data Acquisition System}
A Raspberry Pi 4 was selected for data collection due to its processing capabilities and interface options suitable for handling multiple sensors and devices with varying communication protocols.

\begin{comment}
\begin{figure}[h] 
\centering 
\includegraphics[width=0.45\textwidth]{figures/comm.png} 
\caption{Block diagram of the sensor setup. Yellow highlights indicate the measurement quantities; green highlights indicate the communication types.} 
\label{SensorDiagram} 
\end{figure}
\end{comment}

The sensors were connected as follows:

\begin{itemize} 
\item \textbf{I\textsuperscript{2}C Devices}: The SHT31 temperature and humidity sensor was connected via the Raspberry Pi's built-in I\textsuperscript{2}C bus with integrated pull-up resistors. 
\item \textbf{UART Devices}: The CM1106 CO\textsubscript{2} sensor was connected using the Raspberry Pi's UART interface. 
\item \textbf{USB Devices}: The LCR6100 for impedence measurements and Keysight 34970A for thermocouple data acquisition were connected to the Raspberry Pi 4 with an RS232 to serial converter.
\item \textbf{Analog Devices}: 2 of the I/O pins on Raspberry Pi 4 were configured to measure voltage from a voltage divider circuit for DC resistance, and a SEN0161 pH measurement modules' amplifier circuit.
\item \textbf{Power Requirements}: A separate 5V switch-mode power supply adapter was used to provide voltage to the analog circuitry to eliminate any power fluctuations coming from the Raspberry Pi 4. The digital sensors were powered by the on-board 3.3V regulator on the Raspberry Pi 4. The Raspberry Pi 4 itself was energized with its original 5V 3A power adapter.
\end{itemize}

\begin{figure}[h] 
\centering 
\includegraphics[width=0.47\textwidth]{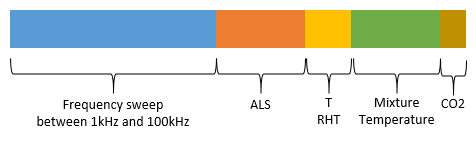} 
\caption{Structure of the data package sampled every minute during fermentation. Optical data is not present here since it required a separate data collection setup.} 
\label{dataPackage} 
\end{figure}

Custom Python scripts were developed to interface with the sensors and manage data acquisition. The collected data were stored in CSV format for subsequent analysis. The structure of the merged data packet, sampled every minute, is shown in Figure \ref{dataPackage}.

\subsection{Experimental Procedure}

The experimental procedure involved several key steps:

\begin{enumerate} 
\item \textbf{Preparation}: Similar combinations of milk and yogurt culture were prepared to ensure consistency across all fermentation runs. The culture was thinned and mixed with 50 ml of milk and then mixed with the entire batch of milk. The mixture was preheated to 43\degree C before it was placed in the oven.

\item \textbf{Sensor Calibration}: All sensors were calibrated according to the manufacturer's specifications prior to testing to ensure accurate measurements. 

\item \textbf{Assembly}: The sensors were integrated into the fermentation container, as depicted in Figures \ref{figLid} and \ref{ImpProbes}. 

\item \textbf{Fermentation Runs}: The preheated yogurt mixture was placed in the container and fermentation was started under controlled conditions. 

\item \textbf{Data Collection}: Sensor data were collected continuously using the Raspberry Pi 4 system throughout the fermentation process. The data were appended to a comma-separated value file for further processing.

\item \textbf{Data Analysis}: The collected data were analyzed to assess the effectiveness of each measurement method in tracking the fermentation process and correlating with pH changes. 
\end{enumerate}

Each sensor system is detailed in subsequent subsections, including methodology, calibration procedures, and specific settings used during the experiments.

\subsection{Sensor Systems and Measurement Methods}

The following sensor systems were evaluated for their effectiveness in monitoring yogurt fermentation.

\begin{itemize} 
\item \textbf{Electrical Impedance and DC Resistance Measurements}: To assess the correlation between electrical properties and pH levels. 
\item \textbf{pH Measurement}: Direct measurement of acidity using pH probes. 
\item \textbf{Optical Permeability}: Monitoring changes in light transmission through the yogurt mixture. 
\item \textbf{CO\textsubscript{2} Concentration}: Measuring CO\textsubscript{2} levels as an indirect indicator of microbial activity. 
\item \textbf{Ambient Temperature and Humidity}: Recording of environmental conditions within the fermentation container. 
\end{itemize}

Each method was selected based on its potential to provide insight into the fermentation process, directly or indirectly related to pH changes and overall yogurt quality.

\subsubsection{pH measurements}
As a result of lactic acid fermentation, yogurt is expected to reach a pH of around 4.3 when the process is complete \cite{shahbook}. For testing purposes, we needed to measure pH continuously for nearly 6 hours and record the data. In addition to the setup shown in Figure \ref{figLid}, A SEN0161 pH measurement kit was connected to the Raspberry Pi 4 via the analog pins. The probe was calibrated using two-point calibration within the acidic range. 
Yogurt quality is directly related to how well the culture is mixed and the pH can vary along the height and width of the container if temperature is uneven \cite{homogen}. For this reason, the fermentation containers were isolated from the base of the oven with a separator to avoid high temperatures at the base.

\begin{figure}[h]
\centering
\includegraphics[width=0.5\textwidth]{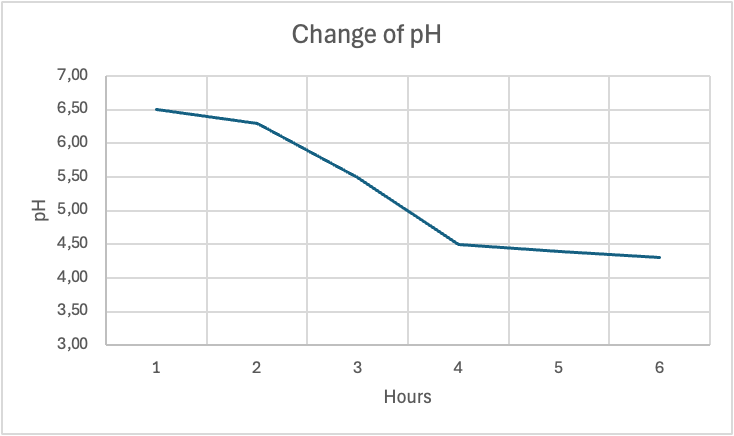}
\caption{Change of pH throughout the fermentation process where the x axis is seconds}
\label{fig:phcurve}
\end{figure}

When Figure \ref{fig:phcurve} is inspected, the desired pH of 4.5 is achieved after 5 hours. 3 liters of Whole cow's milk and 100 mL of regular yogurt culture in liquid form was used. The components were mixed in a common container thoroughly.
\subsubsection{Ambient temperature and humidity measurements}
\begin{figure}[h]
\centering
\includegraphics[width=0.5\textwidth]{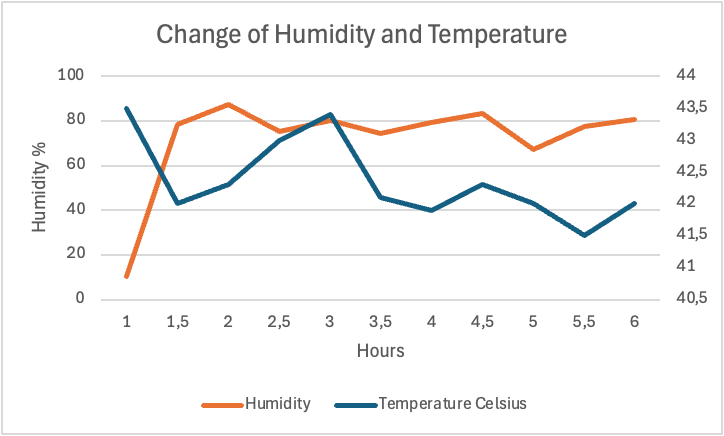}
\caption{The change of relative humidity and temperature inside the fermentation cavity throughout the fermentation process}
\label{fig:temphumidity}
\end{figure}

To track the relative humidity and temperature inside the fermentation cavity, SHT31, which is a digital sensor is used. Our motivation was to observe a definitive trend in both or either of these quantities throughout the fermentation. For the test setup provided in Figure \ref{fig:aligator}, the relative humidity and temperature inside the fermentation cavity is given in Figure \ref{fig:temphumidity}. 

One can observe that when the process starts and the lid is closed, both the temperature and relative humidity start increasing. However, oscillations in relative humidity are considerably higher than in temperature. This is due to the fact that our enclosure is not perfectly sealed and humidity can build up and discharge into the oven cavity periodically. The initial overshoot in the temperature curve is due to the oven's thermal inertia. 

\subsubsection{CO\textsubscript{2} measurements}

\begin{figure}[h]
\centering
\includegraphics[width=0.5\textwidth]{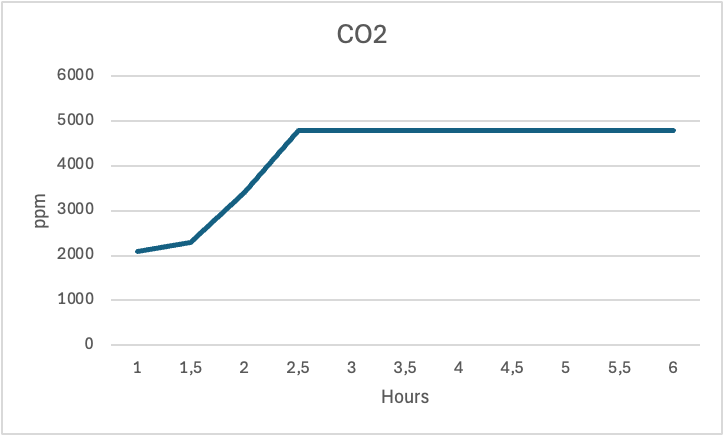}
\caption{The change of CO\textsubscript{2} gas inside the fermentation cavity throughout the process}
\label{fig:co2graph}
\end{figure}

Naturally, \emph{Lactobacillus acidophilus} makes anaerobic respiration, which means it does not use oxygen. As a result, no CO\textsubscript{2} is produced \cite{lactobascillus}. However, the activity of \emph{Streptococcus thermophilus} also causes amino acid and acetaldehyde production, which also produces CO\textsubscript{2} as a side product. \emph{Lactobacillus acidophilus} can then operate more efficiently in a symbiotic manner, accelerating the fermentation process \cite{CHANDAN201731}. Our motivation was to observe the CO\textsubscript{2} concentration inside the fermentation cavity that could somehow be an indicative measure of how the fermentation was proceeding. 
We have used the model CM1106 from Wuhan Cubic which can measure CO\textsubscript{2} concentrations up to 5000 ppm. 
When the results in Figure \ref{fig:co2graph} are inspected, it is obvious that the cavity was saturated with CO\textsubscript{2} gas and the sensor signal was clipped. This is because the CO\textsubscript{2} concentration in the cavity was much greater than 5000 ppm. Experiments were repeated using a 10000 ppm sensor, but the measurements were saturated just like the first one. 

\subsubsection{Optical permeability measurements}
The literature provides decent findings regarding optical methods to track milk and fermentation state \cite{opticalfat}\cite{laserdiffraction}. For this reason, we further inspected this issue, but with a simpler, more fundamental setup. The rectangular cavity for this test was designed in Siemens NX and was 3D printed using SLA material. For the optically transparent parts, pieces of clear plexiglass were used. For the light source, A 1W power LED with a wavelength of 400nm was picked. For the detector, a regular light-dependent resistor (LDR) was not preferred due to their temperature dependence, low resolution and drifting issues. Instead, Si1133, which is originally a UV spectrometer with built-in ambient light sensing capabilities was used. It features a resolution of less than 100mlx, and a spectra of 128klx, making it a viable choice for low and intense light applications. The light source and sensor were fixed to the setup with black tape to cover all possible openings where external lighting can interfere with the results.

\begin{figure}[h]
\centering
\includegraphics[width=0.5\textwidth]{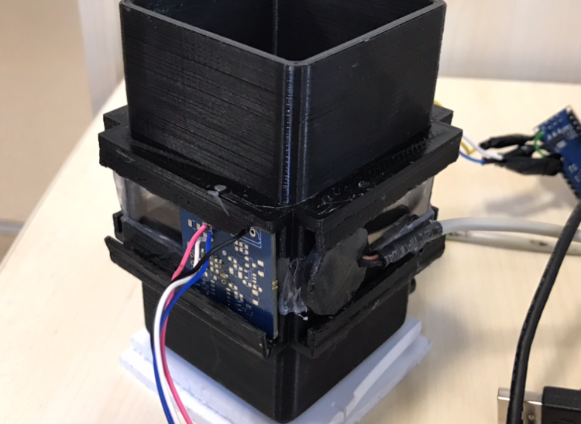}
\caption{Setup for measuring the optical permeability of the yogurt mixture}
\label{fig:olayy}
\end{figure}

In Figure \ref{fig:olayy}, the light source and the ALS sensor were oriented 90° with respect to each other, allowing for the yogurt mixture to fill in between. 

\begin{figure}[h]
\centering
\includegraphics[width=0.5\textwidth]{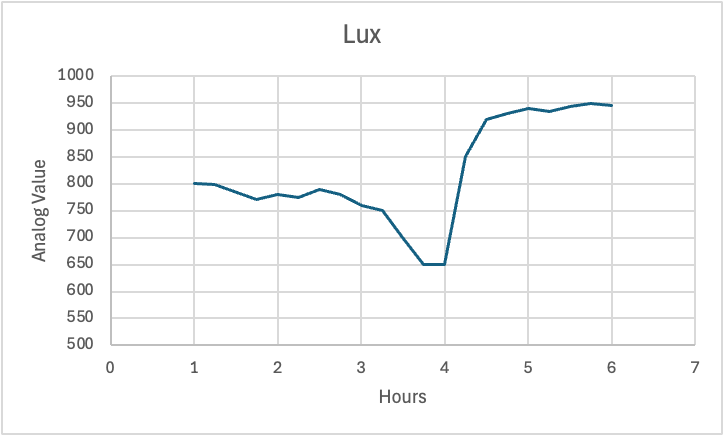}
\caption{The change of optical permeability of the yogurt mixture throughout the fermentation process}
\label{fig:optigrafik}
\end{figure}

When the results in Figure \ref{fig:optigrafik} were inspected, it was obvious that the optical permeability of the yogurt mixture follows a sudden, increasing trend as the coagulation starts. Although this coagulation was not quantized with measuring equipments, the lid was opened and the surface of the yogurt was inspected visually for coagulation.

%\subsection{Imaging the surface of the yogurt}
%We wanted to inspect the surface of the yogurt mixture throughout the fermentation process for any indicative visuals. A Logitech C270 was connected to the Raspberry Pi using the USB 3.0 port. It was defined as a serial device and told to capture an image every minute. Because the inside of the cavity was dark, A spectrometer with built-in LED's were also connected from the USB 3.0 port, allowing us to switch on the white LED's for better image quality. In Figure \ref{fig:kapak}, the webcam is located at the center of the pink lid, facing downwards.

\subsubsection{DC resistance measurements}
It is possible to track the conductivity and resistance of various dairy products for analysis \cite{condmilk}. These analysis may often include spoilage detection or fat content determination. However, this method should be taken with a grain of salt because conductive liquids can undergo electrolysis when a voltage difference is present with conductive plates. Commercially available conductivity measurement probes utilize low voltage levels and inert coatings to overcome this issue. 

\begin{comment}
  \begin{figure}[h]
\centering
\includegraphics[width=0.5\textwidth]{figures/pins.jpg}
\caption{Conductive headers for electrical measurements}
\label{fig:pinss}
\end{figure}  
\end{comment}

As our measurement setup, we have utilized an ordinary voltage divider circuit to observe the resistance of the yogurt mixture. The voltage level was fed to the Raspberyy Pi 4, and the analog signal was converted to a voltage level, then to a resistance value. However, due to the DC voltage difference, the anode pin quickly dissolved into the yogurt, making the results unstable. To overcome this, we have used pure platinum electrodes, both for the anode and cathode. When the experiments were repeated, there were no damage to the platinum contacts. However, the yogurt between the electrodes were polarized, meaning that the positive and negative ions were attracted to the anode and cathode, causing the resistance reading to increase continuously. This was confirmed when the yogurt in between the electrodes was inspected visually. 
\subsubsection{Impedance measurements}
As previously mentioned, the literature focuses extensively on bio impedance measurements to either detect living organisms or perform quality check \cite{bioimpmatlab}\cite{bioimpcells}\cite{impmatlab}\cite{condic}. Our first step was to replicate the results that were already present in the literature. Figure \ref{ImpProbes} shows the attachment of the alligator clips to the electrical contacts. It is possible to analyze bovine milk based on their EIS, by sweeping a frequency range of 1kHz to 100kHz \cite{cheesestarters}. Based on this information, the LCR meter was set up to scan this interval with 1kHz steps. The measurements involved capturing the total impedance's magnitude, denoted as $|Z|$. To configure the device, specific external text commands were needed, and the RS232 protocol was employed for sending and receiving data. A tailored Python code was created and executed on the Raspberry Pi 4, which was linked to the LCR meter. The measurements were collected continuously throughout the entire fermentation process. To create a secure seal between the probes and the container, a two-component epoxy adhesive was utilized. To prevent any contamination of the yogurt mixture from the epoxy, the adhesive was exclusively applied to the external surface of the container. Following the completion of each measurement cycle by the LCR meter, it transmits the data as a bundle, and the Raspberry Pi 4 logs each line into a .csv file. 

\begin{figure}[h]
\centering
\includegraphics[width=0.5\textwidth]{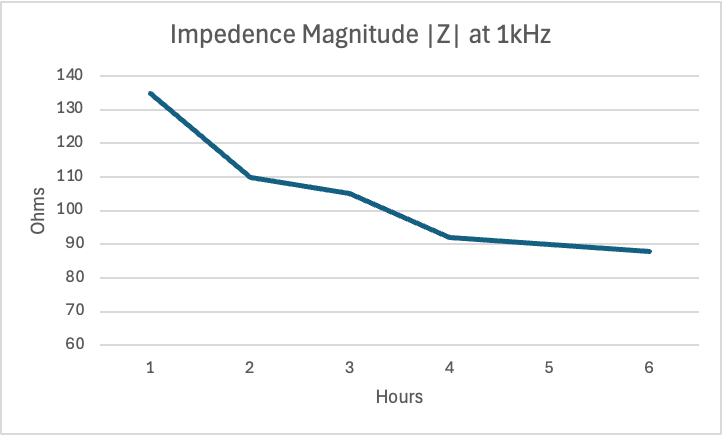}
\caption{Change of the absolute value of total impedance for 5, 10 and 15kHz}
\label{fig:impgraph}
\end{figure}

Figure \ref{fig:impgraph} illustrates the variations in $|Z|$ during the fermentation process. Frequencies up to 100kHz were examined, but they exhibited no significant deviations from the values displayed here. To maintain clarity, these additional frequencies are omitted from the graph. Nevertheless, the results for all three frequencies had negligible differences.

Following 25 tests, the probes in direct contact with the yogurt (excluding the alligator clips) displayed minor discoloration, with no substantial signs of corrosion or rust. The probes were not coated with an inert element, in contrast to DC resistance measurement probes, which were replaced with fresh ones.

\begin{figure}[h]
\centering
\includegraphics[width=0.49\textwidth]{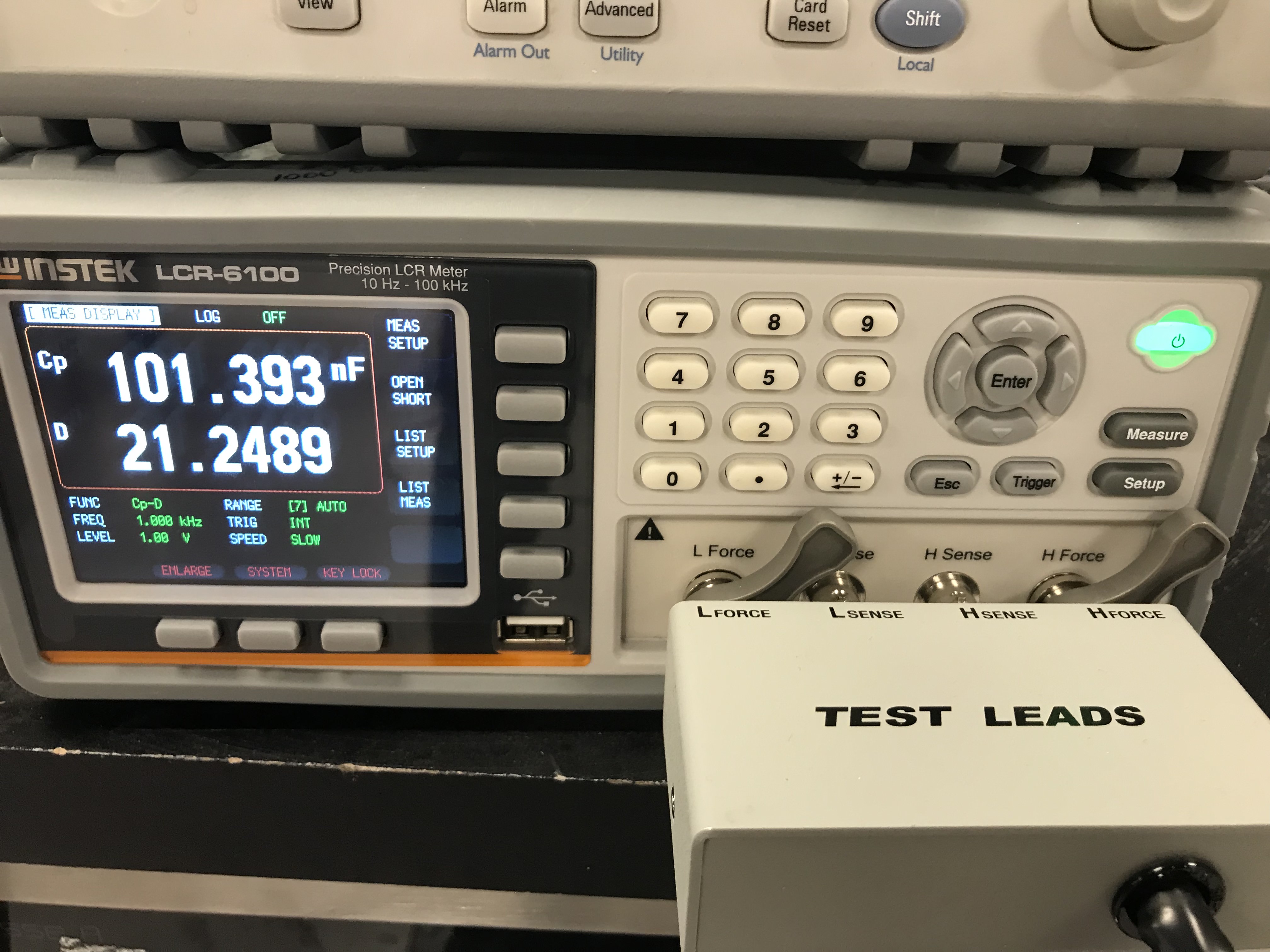}
\caption{LCR6100 is used for measuring the absolute impedance Z of yogurt}
\label{fig:pinss}
\end{figure}

\section{Evaluation of the results}
In this section, we will assess the experiments conducted in the preceding section and analyze their significance and connection to our objective. Ultimately, we will discuss their strengths and weaknesses to provide o thorough evaluation.

\subsubsection{pH measurement results}

\textbf{Pros}
\begin{itemize}
  \item \textbf{Direct Indicator of Sourness:} pH measurements provide a direct representation of the sourness level of yogurt, a critical quality factor.
  \item \textbf{Industry Standard:} pH measurement is widely used in the food industry, facilitating comparison with established practices.
\end{itemize}

\textbf{Cons}
\begin{itemize}
  \item \textbf{Calibration Requirement:} pH meters require regular calibration, adding complexity and maintenance.
  \item \textbf{Cost:} High-quality pH meters can be relatively expensive, limiting accessibility.
  \item \textbf{Maintenance:} Proper cleaning and storage are essential to prevent drift and ensure accurate measurements.
  \item \textbf{Single Parameter:} pH measurements provide acidity data but lack information on other fermentation aspects.
  \item \textbf{Electrode Sensitivity:} pH electrode sensitivity to contamination or damage can affect results.
  \item \textbf{Contact:} Needs a direct contact (intrusive) to the yogurt mixture in order to measure, which might be undesirable in certain cases.
\end{itemize}

While pH probes offer direct and accurate means of assessing the sourness level in yogurt, their high cost and calibration requirements make them less practical for certain cases such as widespread use in small-scale or home-based yogurt production. Combined with the fact that yogurt has a gel-like structure, cleaning and maintaining pH probes becomes a challenge. These limitations underscore the need to explore alternative, cost-effective sensor methods to enable more accessible and precise yogurt fermentation monitoring.

\subsubsection{Ambient temperature and humidity measurement results}

\textbf{Pros}
\begin{itemize}
  \item \textbf{Closeness to Milk Temperature:} Ambient temperature measurements inside the yogurt fermentation chamber closely approximate the milk's temperature, which is relevant to fermentation.
  \item \textbf{Non-Invasive:} Measuring ambient conditions is non-invasive and does not directly interact with the yogurt, making it a convenient and passive monitoring approach.
\end{itemize}

\textbf{Cons}
\begin{itemize}
  \item \textbf{Limited Fermentation Insight:} While temperature and humidity measurements can provide data on the environment, they offer limited insights into the actual progress of yogurt fermentation.
  \item \textbf{Dependence on External Factors:} Temperature and humidity within the fermentation chamber can be influenced by external factors and may not accurately represent the yogurt's internal conditions.
\end{itemize}

While measuring ambient temperature and humidity within the fermentation chamber may seem initially promising due to their proximity to milk temperature, these measurements ultimately prove to be ineffective for reliably tracking the yogurt fermentation process. Their limited insight into the actual fermentation progression, susceptibility to external influences, and lack of specificity regarding critical fermentation parameters render them an unreliable option for comprehensive yogurt fermentation monitoring. However, strong monitoring systems can be made if ambient temperature and humidity sensing is coupled with other methods such as pH monitoring.

\subsubsection{CO\textsubscript{2} measurement results}

\textbf{Pros}
\begin{itemize}
  \item \textbf{Indirect Indicator of Bacterial Activity:} CO\textsubscript{2} measurements can serve as an indirect indicator of bacterial activity during yogurt fermentation since certain bacteria produce CO\textsubscript{2} as a byproduct\cite{shahbook}.
  \item \textbf{Non-Invasive:} Measuring CO\textsubscript{2} is non-invasive and does not disturb the yogurt or the fermentation process.
\end{itemize}

\textbf{Cons}
\begin{itemize}
  \item \textbf{Sensor Saturation:} In a closed fermentation enclosure, CO\textsubscript{2} levels quickly saturated the sensor, reaching peak values (e.g., 5000ppm) and rendering the sensor ineffective. Even higher-range sensors (e.g., 10000ppm) were similarly saturated.
  \item \textbf{Price:} CO\textsubscript{2} sensors can be relatively expensive, especially those with wider detection ranges, making them cost-prohibitive for some applications.
  \item \textbf{Limited Added Value:} Due to sensor saturation and the limited additional insights gained, the use of CO\textsubscript{2} measurements did not significantly enhance the yogurt fermentation monitoring process in this context.
  \item \textbf{Inability to Differentiate:} CO\textsubscript{2} measurements, on their own, do not differentiate between different bacterial strains or provide information on other key fermentation parameters like pH or texture. 
\end{itemize}
While CO\textsubscript{2} measurements offer the advantage of serving as an indirect indicator of bacterial activity and validating the hypothesis regarding bacterial quantity, their effectiveness is hindered in closed fermentation enclosures where sensor saturation occurs, even at 10000ppm. Intrinsically, most end users do not cover their fermentation cups entirely to let the excess moisture escape for a thicker yogurt. Coupled with the relatively high cost of these sensors and their limited additional value in enhancing yogurt fermentation monitoring, their application is not deemed a practical choice for comprehensive and cost-effective tracking of the fermentation process.

\subsubsection{Optical permeability measurement results}

\textbf{Pros}
\begin{itemize}
  \item \textbf{Texture Indicator:} Optical permeability measurements offer valuable insights into yogurt texture as they correlate with the coagulation process. An increase in light permeability corresponds to changes in yogurt consistency.
  \item \textbf{Non-Invasive:} This method is non-invasive, meaning it doesn't physically interact with the yogurt, making it a convenient and passive approach for monitoring.
  \item \textbf{Definitive Information:} Optical permeability provides a definitive indicator of the coagulation process, allowing for real-time monitoring of yogurt's transformation.
\end{itemize}

\textbf{Cons}
\begin{itemize}
  \item \textbf{Mechanical Constraints:} A potential drawback is the need for precise mechanical positioning of the sensor and the white LED, which can be challenging due to mechanical constraints and moving parts. This may introduce complexity to the setup.
  \item \textbf{Lack of pH Information:} Optical permeability measurements do not provide data on pH levels, a critical factor in yogurt quality. Therefore, this method alone may not capture all relevant parameters.
  \item \textbf{Limited to Texture:} While valuable for texture assessment, optical permeability measurements do not offer insights into other fermentation aspects, such as flavor development.
\end{itemize}

\subsubsection{DC resistance measurement results}

\textbf{Pros}
\begin{itemize}
  \item \textbf{Strong connection to acidity:}  The pH level exhibits a direct correlation with electrical resistance \cite{ecandph}. Likewise, monitoring the DC resistance of the yogurt mixture throughout the fermentation process provides a potential avenue to establish a correlation with pH.
  \item \textbf{Economical solution:} 
 Assembling a resistance measurement circuit is cost-effective, comparable to crafting a voltage divider circuit, and only necessitates passive electronic components.
\end{itemize}

\textbf{Cons}
\begin{itemize}
  \item \textbf{Electrochemical reaction:} 
If two electrodes are immersed in a conductive solution and subjected to a direct current voltage, the anode will undergo oxidation and may melt, ultimately dispersing into the solution \cite{electrolysis}. However, in yogurt production, this scenario is deemed unacceptable due to food safety regulations \cite{foodregulation}.
  \item \textbf{Polarization:} If sufficient exposure to DC voltage is sustained, electrical polarization may occur \cite{polarization}. This causes charges to migrate towards the electrical contacts, increasing the resistance of the medium. In the case of yogurt, it was observed that polarization not only causes resistance increase, but also phase change in the yogurt structure, therefore creating an undesirable situation.
  \item \textbf{Temperature dependent:} Although electrical resistance is directly proportional with temperature for solids, it is inversely proportional for liquids \cite {inverse}. Temperature control or compensation is necessary to obtain accurate resistance measurements.
\end{itemize}

In conclusion, while DC resistance measurement offers several advantages such as its strong correlation with acidity and cost-effectiveness, certain drawbacks need to be addressed. Electrochemical reactions resulting in electrode oxidation and potential contamination of the solution pose significant concerns, particularly in the context of food safety regulations. Additionally, issues like electrical polarization and temperature dependence can impact the accuracy and reliability of resistance measurements during yogurt fermentation. Addressing these challenges through proper electrode selection, monitoring, and temperature control measures will be crucial for ensuring the efficacy and safety of DC resistance-based monitoring systems in yogurt production processes.

\subsubsection{Impedance measurement results}

\textbf{Pros}
\begin{itemize}
  \item \textbf{Measurement flexibility:} For dairy products, electrical impedance spectroscopy allows for deeper feature extraction \cite{milkimp}\cite{milkcharactrization}. This was also confirmed by our experiments. However, we did not observe any major trend differences during frequency sweeping.
  \item \textbf{Strong connection to acidity: }The pH level demonstrates a direct relationship with electrical impedance. Similarly, tracking the impedance of the yogurt mixture during fermentation offers a possible means to establish a connection with pH levels.  
  \item \textbf{Minimal electrochemical reaction:} Since electrical impedance is measured with alternating currents, the chemical disturbance is distributed among the two contact points. Combined with the fact that utilization of low voltages for impedance measurement is possible without significant amplification circuitry, chemical concerns are less for electrical impedance measurements.
\end{itemize}

\textbf{Cons}
\begin{itemize}
  \item \textbf{Temperature dependent:} While electrical resistance increases with temperature for solids, it decreases for liquids \cite{inverse}. Ensuring accurate resistance measurements requires temperature control or compensation mechanisms.
  \item \textbf{Potentially high cost:} Impedance measurement circuits or integrated circuits might be costly. Combined that they need a host device to receive various operating commands, both the measurement circuit and its peripherals may contribute to high cost.

Even though potentially high costs and temperature dependency issues seem significant for electrical impedance measurement, its benefits majorly outweigh its drawbacks. The validity of electrical impedance measurement to track yogurt fermentation has proven itself by both our experiments and the literature. It is imperative to note that fine tuning the impedance measurement circuitry or coming up with cost effective impedance measurement alternatives is beyond the scope of this paper. The sole purpose of this paper is to present an array of sensor options to track yogurt fermentation. It is up to the readers' appreciation to use or utilize any of the presented methods.

\end{itemize}

\begin{comment}
    \begin{figure}
    \centering
    \includegraphics[width=0.5\linewidth]{figures/yoghurt_probe2.png}
    \caption{Custom electrical impedance measuring probe used in the yogurt machine.}
    \label{fig:yogurt_probe}
\end{figure}

\begin{figure}
    \centering
    \includegraphics[width=0.75\linewidth]{figures/ym2.png}
    \caption{Final version of the yogurt machine developed for mass production.}
    \label{fig:yogurt_machine}
\end{figure}

\begin{figure}
    \centering
    \includegraphics[width=1\linewidth]{figures/DualPlottTempImp.png}
    \caption{\textcolor{blue}{Change of impedance and temperature during fermentation.}}
    \label{fig:yogurt_impgraph}
\end{figure}
\end{comment}

%\section{Findings}
%Findings ayri bir section olmali mi?

\section{Discussion}
This study has examined various sensor technologies to monitor the fermentation process of yogurt, an increasingly popular fermented dairy product. Our contributions are particularly significant in providing a comprehensive evaluation of these technologies in a domestic setting, which has not been extensively explored in existing literature.
 
The primary contribution of this research lies in the detailed examination of sensor technologies such as pH measurement, ambient temperature and humidity monitoring, CO\textsubscript{2} detection, optical permeability, and electrical properties measurement. Each method's effectiveness in capturing the critical parameters of the yogurt fermentation process was thoroughly assessed. Our findings suggest that while traditional pH measurement offers the most direct indicator of fermentation progress through acidity levels, it may not be the most practical or cost-effective solution for home-based yogurt production due to its maintenance and calibration requirements.
 
Temperature and humidity measurements, although non-invasive and closely related to the fermentation environment, were found to offer limited insights into the fermentation process itself. CO\textsubscript{2} monitoring presented an indirect measure of bacterial activity but faced limitations due to sensor saturation in closed environments. Optical permeability and electrical impedance spectroscopy emerged as promising non-invasive techniques, providing valuable information on the yogurt's textural changes and acidity levels, respectively.
 
The study’s novel approach to utilizing these sensor technologies in a home-based yogurt fermentation context aims to bridge the gap between professional and domestic production. It addresses the growing consumer interest in homemade yogurt, driven by concerns over mass-produced food's safety and ingredient transparency.
 
Looking forward, the potential for integrating these sensors into a consumer-friendly yogurt making appliance poses an exciting avenue for further research. Such integration could democratize the ability to produce high-quality yogurt at home, providing consumers with real-time data to optimize the fermentation process according to personal taste and nutritional preferences.

\section{Conclusion}

In this work, we conducted a systematic exploration and experimental assessment of multiple sensor technologies to monitor the yogurt fermentation process. Our investigation covered direct (e.g., pH probes) and indirect (e.g., electrical impedance, optical permeability, CO\textsubscript{2} concentration, ambient conditions) methods, evaluated both individually and in combination. While the findings are based on yogurt fermentation, the methodological approach and comparative insights can be readily extended to other fermented food products or analogous industrial bioprocesses.

From a broader perspective, this study contributes new knowledge about sensor placement, calibration, and integration—factors essential for designing robust systems that balance affordability with accuracy. Our work highlights opportunities to combine different measurements, potentially harnessing data fusion and machine learning to increase reliability in fermentation monitoring. For instance, pairing electrical impedance with optical methods or CO\textsubscript{2} sensing could yield a more holistic view, enabling faster industrial throughput, reduced resource usage, and improved product consistency.

In terms of future directions, these insights can be extended to fermentation processes in biotechnology and pharmaceuticals where cell growth and metabolic production require similarly fine-grained, non-destructive monitoring. Researchers may adapt the present methodology—frequency sweeping in impedance measurements, multipoint optical sensing, or multi-sensor data fusion—to better capture the intricate chemical and physical transformations that arise in complex fermentation workflows. The modularity and scalability of the approaches highlighted here further underscore their potential for customized or large-scale system integration.

Ultimately, this work underscores that mapping out the full landscape of sensor modalities—beyond just pH—can drive innovation in the design of fermentation monitoring solutions for both large-scale industry and specialized research endeavors. By offering a cross-comparison of sensing technologies, we aim to spark continued refinement and expansion of real-time process control in the broader field of fermentation science and engineering.

\bibliography{yoghurt-jsen}{}
\bibliographystyle{IEEEtran}

%\begin{IEEEbiography}[{\includegraphics[width=1in,height=1.25in,clip,keepaspectratio]{figures/bio/a2.jpeg}}]{Ege Keskin} has Bahchelor's and Master's degree in Mechatronics Engineering from Bahçeşehir University. His current research involves deep learning and its adaptation to gesture recognition systems.
%\end{IEEEbiography}

\begin{IEEEbiography}[{\includegraphics[width=1in,height=1.25in,clip,keepaspectratio]{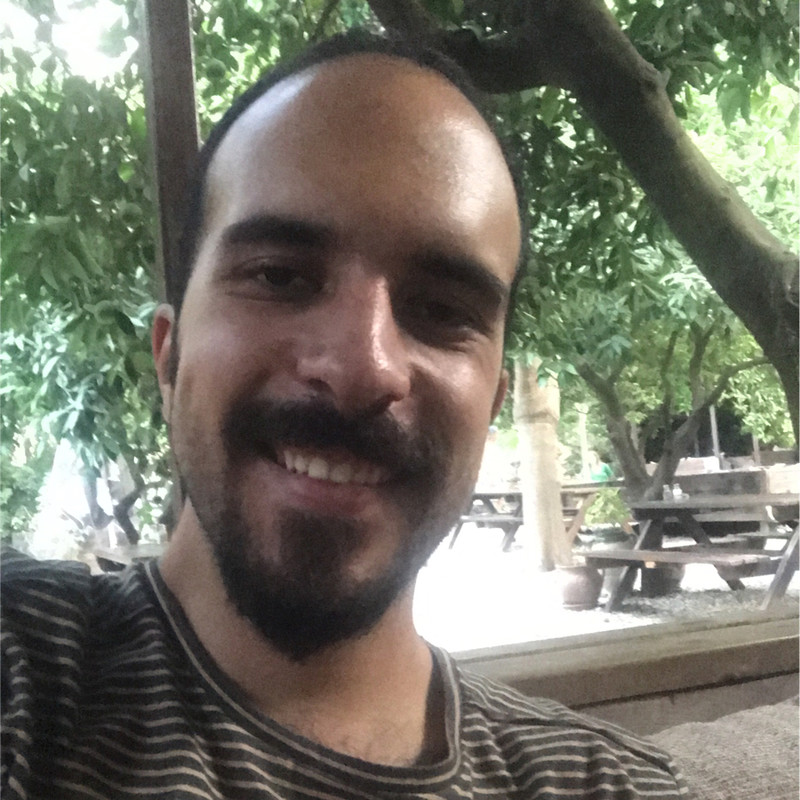}}]{Ege Keskin} received his Bachelor's at 2018 and Master's degrees at 2020 in Mechatronics Engineering from Bahçeşehir University, Istanbul, Türkiye. He is currently pursuing a Ph.D. degree in Design, Technology, and Society at Koç University - Arçelik Research Center for Creative Industries, Istanbul, Türkiye.

Ege Keskin also currently serves as a Senior Specialist Engineer in the Sensor Technologies Department at Beko Corporate R\&D Directorate, Çayırova, Istanbul, Türkiye. His primary research interests lie in the development of sensor technologies for domestic appliances. Moreover, he is actively engaged in exploring the applications of deep learning techniques, particularly in the field of gesture recognition systems.
\end{IEEEbiography}

\begin{IEEEbiography}[{\includegraphics[width=1in,height=1.25in,clip,keepaspectratio]{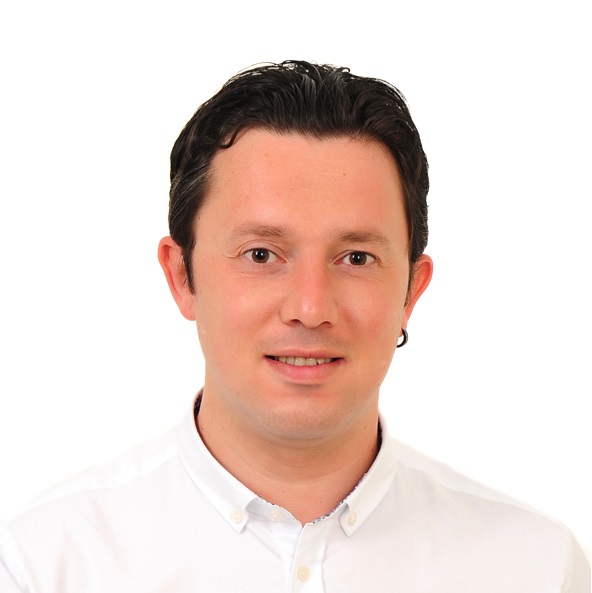}}]
{İhsan Ozan Yıldırım}
(Student Member, IEEE) received his B.Sc. degree in Electronics and Communications Engineering from Yıldız Technical University in Istanbul, Türkiye, in June 2010. He obtained his M.Sc. degree from the Electrical and Electronics Engineering Department at Koç University,Istanbul, Türkiye, in January 2013. Currently, he is actively pursuing a Ph.D. degree in Design, Technology, and Society at Koç University - Arçelik Research Center for Creative Industries, İstanbul, Türkiye.

Presently serving as a Senior Lead Engineer in the Sensor Technologies Department at Beko Corporate R\&D Directorate, Çayırova, İstanbul, Türkiye. His research focus revolves around smart device prototyping, sensor system development, air quality sensing, and data science applications related to sensor technologies in consumer electronics.
\end{IEEEbiography}

\end{document}